\documentstyle[times,graphics,astrobib,amssymb]{mn2e}

\newcommand{\be}{\begin{equation}}
\newcommand{\e}{\end{equation}}
\newcommand{\bear}{\begin{eqnarray}}
\newcommand{\ear}{\end{eqnarray}}

\newcommand{\f}{\frac}
\newcommand{\de}{{\rm d}}

\begin{document}

\title[Reionization scenarios]
{Updating reionization scenarios after recent data}
\author[Choudhury \& Ferrara]
{T. Roy Choudhury$^{1}$\thanks{E-mail: tirth@cts.iitkgp.ernet.in}~
and
A. Ferrara$^{2}$\thanks{E-mail: ferrara@sissa.it}\\
$^{1}$Centre for Theoretical Studies, Indian Institute of Technology,
Kharagpur 721302, India\\
$^{2}$SISSA/ISAS, via Beirut 2-4, 34014 Trieste, Italy}

\maketitle

\date{\today}

\begin{abstract}
The recent release of data on (i) high redshift source counts from NICMOS HUDF,
and (ii) electron scattering optical depth from 3-year WMAP, require a re-examination of 
reionization scenarios. Using an improved self-consistent model, based on Choudhury \& 
Ferrara (2005), we determine the range of reionization histories which can match a wide 
variety of data sets simultaneously.
From this improved analysis we find  that hydrogen reionization starts 
around $z = 15$, driven by the metal-free stars (with normal Salpeter-like IMF),
and is 90\% complete by $z \approx 10$. The photoionizing power of PopIII stars fades 
for $z \lesssim 10$ because of the concomitant action of radiative and chemical feedbacks, which
causes the reionization process to stretch considerably and to end only by $z \approx 6$. 
The combination of different data sets still favours a non-zero contribution from 
metal-free stars, forming with efficiencies $> 2$\%.           
\end{abstract}
\section{Introduction}

The determination of the high Thomson electron scattering optical 
depth $\tau_{\rm el} = 0.17 \pm 0.04$ in the WMAP 1st year data 
\cite{ksb++03,svp++03} had been a subject of extensive theoretical
study over the last few years. 
For sudden reionization models, the high value of $\tau_{\rm el}$ 
implied that reionization
occurred at very high redshifts $z \approx 15$.
This scenario seemed to be at tension with the 
measured Gunn-Peterson (GP) optical depth $\tau_{\rm GP}(z=6) \gtrsim 6$ 
from the absorption line experiments
of $z \gtrsim 6$  Sloan
Digital Sky Survey (SDSS) QSOs \cite{fnl+01,fss++03,fsb++05}.
Such high values of $\tau_{\rm GP}$ seem to indicate that 
reionization was complete only at $z \approx 6$.

Using a self-consistent formalism confronted with a wide range of observational data sets 
(redshift evolution of Lyman-limit absorption systems,
GP and electron scattering optical depths, 
temperature of the mean intergalactic gas,  
and cosmic star formation history), 
\nocite{cf05} Choudhury \& Ferrara (2005; hereafter CF05)
showed that the most favourable model 
is the one in which hydrogen reionization was complete
at $z \approx 12$. 
By using the statistics of dark gaps in the Ly$\alpha$ transmitted flux, 
this early reionization model was also shown not to be in conflict 
with QSO absorption line experiments at $z \gtrsim 6$ \cite{gcf05}.
However, 
a nearly equally good fit to the data could be achieved
for late reionization scenarios too (i.e., scenarios
in which reionization is complete only at $z \approx 6$), particularly
if one relaxes the constraints on $\tau_{\rm el}$. Given this, 
the need was to find an additional way to distinguish between the early and
late reionization models,
either through new
theoretical methods \cite{gcf05} or from additional observational constraints.

Fortunately, two new sets of data have been made available
recently which could help in constraining the reionization history. First
is the observations of high redshift sources
in the NICMOS HUDF \cite{bitf05}, where the analysis indicate that the number
of sources at $z \approx 10$ should be three or fewer. 
This inevitably rules out the occurrence of very massive
($\gtrsim 300 M_{\odot}$) stars \cite{sf05}.
The second set
of observations is the release of 3-year WMAP data \cite{hnb++06,phk++06,sbd++06}, which gives a 
lower value of $\tau_{\rm el} = 0.09 \pm 0.03$, thus
questioning very early reionization scenarios.

Given these new data sets, it is important to find out the
updated constraints on reionization using self-consistent
models. More importantly, one has to address the issue as to which
reionization histories
are still viable and which sources are responsible for it.
In this work, we extend the model of CF05 incorporating 
some additional physics (like chemical feedback) thus reducing the 
number of free parameters. We
then confront the model with a wide variety of available data sets (including the two 
most recent ones above) with the aim of identifying a
set of parameter values which can fit {\it all} the
data sets simultaneously.

\section{Model Description}

In this Section, we first summarize the main features of the model
introduced in CF05; following that we discuss the modifications and 
improvements made for the purpose of this work.

\subsection{Summary of the model}

The main features of the semi-analytical model used in CF05 could
be summarized along the following points (for detailed explanations
see CF05). The model accounts for IGM inhomogeneities by adopting 
a lognormal distribution according to the method outlined in \citeN{mhr00}; 
reionization is said to be complete once all the low-density regions (say, with overdensities $\Delta < 
\Delta_{\rm crit} \sim 60$) are ionized. 
Hence, the distribution of high density regions determines the 
mean free path of photons
\be
\lambda_{\rm mfp}(z) = \f{\lambda_0}{[1 - F_V(z)]^{2/3}}
\label{eq:lambda_0}
\e
where $F_V$ is the volume fraction of ionized regions and $\lambda_0$ is a normalization constant fixed 
by comparing with low redshift observations. We follow the ionization and thermal histories 
of neutral, HII and HeIII regions simultaneously and self-consistently, treating the IGM
as a multi-phase medium. As reionization by UV sources is accompanied by heating of the medium, which 
can suppress star formation in low-mass halos, we compute the corresponding radiative feedback self-consistently
from the evolution of the thermal properties of the IGM.
To calculate the ionizing flux, three sources have been assumed: (i) metal-free (i.e., PopIII) stars, which dominate 
the flux at high redshifts. In CF05, they were assumed to be massive ($M \geq 100 M_{\odot}$);
(ii) PopII stars with sub-solar metallicities having a Salpeter IMF in the mass range $1 - 100 M_{\odot}$. 
The transition redshift, $z_{\rm trans}$ 
from PopIII to PopII stars was a free parameter, usually fixed to be $z_{\rm trans} \gtrsim 10$; (iii) QSOs, which are 
significant source for hard photons at $z \lesssim 6$; however they have no effect on the IGM at higher redshifts.

\subsection{New features}

We now discuss the additional physics we have incorporated in this work in order to further improve the
predictive power of the model.  
\begin{itemize}
\item {\it Radiative feedback}: We have assumed that no haloes with virial temperatures lower than $10^4$ K are able 
to form stars; this completely neglects the contribution of minihaloes, which is now strongly supported by the 3-year 
WMAP data \cite{hb06}.

\item {\it Chemical feedback}: The main limitation of CF05 model was the idealized PopIII $\rightarrow$ PopII transition 
which was assumed to start at $z_{\rm trans}$ and last for a dynamical time of the halo. According to the standard
chemical feedback interpretation \cite{sfno02,sfsob03}, the transition is driven by the enrichment of the medium which 
forces a drastic change in the fragmentation properties of star-forming clouds when metallicity exceeds
the critical value of $Z_{\rm crit} = 10^{-5\pm 1} Z_{\odot}$ \cite{sfno02,sfsob03}.
Such feedback-regulated transition has been studied in detail by \citeN{ssfc05}, using a merger-tree approach to determine 
the termination of PopIII star formation in a given halo. We incorporate the same prescription in our model (using Fig 3 of \citeNP{ssfc05}), 
which allows us to compute the transition in a self-consistent manner. The main difference with respect to CF05 is 
that the transition occurs over a prolonged epoch, i.e., no precise transition redshift can be identified.

\item {\it IMF of PopIII stars}: In CF05, a top-heavy IMF for PopIII stars was used, which was found to be disfavoured 
by a combination of constraints from source counts at $z \approx 10$ 
and the first year WMAP data \cite{ssfc05}. 
In this work 
we use a very ``conservative'' 
assumption that the metal-free PopIII stars have a simple Salpeter IMF, just like the PopII stars, which 
is similar to the hypothesis made in \citeN{cfw03}. One should keep
in mind that the recent 3-year WMAP data need not necessarily
rule out the possibility that PopIII stars have a top-heavy IMF; however,
we limit ourselves to the most conservative model and check whether 
it can match all available observations.

\item {\it Escape fraction}: In CF05, the escape fractions for PopII and PopIII stars, $f_{\rm esc, II}$ and 
$f_{\rm esc, III}$, were considered as free parameters, independent of the halo mass, 
$M$, and redshift $z$.  In reality, the situation is quite complex and the escape fractions
do depend on both $M$ and $z$. Unfortunately, there is still no good understanding of the process so as 
to model it theoretically.

In this work, we retain the assumption that the escape fraction is independent of $M$ and $z$; 
however, we use a physical argument to relate the escape fraction of PopII and PopIII stars. 
This is based on the fact that the escape fraction should scale according to the number of ionizing 
photons produced by a given source.  Let $N_{\rm abs}$ denote that number of photons that can be {\it potentially} 
absorbed by the star-forming halo (which can be quite different from the number
of photons actually absorbed).  It is usually proportional to the quantity ${\cal C} \alpha_R(T) n_H n_e$, 
where ${\cal C}$ is the clumping factor of the halo gas density inhomogeneities.
There are further uncertainties related to the distribution of stars within the halo, and we assume 
that such uncertainties can be absorbed within the proportionality factor.
Let $N_{\gamma, {\rm II}}$ ($N_{\gamma, {\rm III}}$) denote the number of photons produced by PopII (PopIII) stars 
per unit mass of star formed and $\epsilon_{*,{\rm II}} (\epsilon_{*,{\rm III}})$ denote the star-forming efficiency of the population.  We can then define the parameter
\be
\eta_{\rm esc} \equiv \f{N_{\rm abs}}{\epsilon_{*,{\rm II}} N_{\gamma, {\rm II}}}
\label{eq:eta_esc}
\e
which measures the fraction of photons absorbed in the halo.
Then one can write the relation
\be
f_{\rm esc, II} = 1 - {\rm Min}\left[1,\f{N_{\rm abs}}{\epsilon_{*,{\rm II}} N_{\gamma, {\rm II}}}\right] 
= 1 - {\rm Min}[1,\eta_{\rm esc}]
\label{eq:f_esc_2}
\e
which takes into account the fact that $f_{\rm esc, II} = 0$ 
if $\eta_{\rm esc} > 1$ (which essentially means that the
halo is capable of absorbing more photons than what is produced
by the stars and thus all the photons produced are
absorbed within the halo). Note that a higher value of 
$\epsilon_{*,{\rm II}}$ would give a higher $f_{\rm esc, II}$
signifying that a higher fraction of photons will escape
if the number of photons produced is larger.

We now make the simplifying assumption that $N_{\rm abs}$ depends 
only on the properties of the halo and is independent
of the nature of the stellar source. Then the escape fraction for 
PopIII stars would be given by
\bear
f_{\rm esc, III} &=& 1 - {\rm Min} \left[1,\f{N_{\rm abs}}{\epsilon_{*,{\rm III}} N_{\gamma, {\rm III}}}\right]\\
&=& 1 - {\rm Min} \left[1,\f{\epsilon_{*,{\rm II}} N_{\gamma, {\rm II}}} {\epsilon_{*,{\rm III}} N_{\gamma, {\rm III}}}
\eta_{\rm esc}\right]
\label{eq:f_esc_3}
\ear
which relates the escape fractions of the two stellar populations. Note that no assumptions about the
gas density structure has been made; we simply used the fact that a higher fraction of photons will escape
if the number of photons produced is larger.
The above prescription can be extended to  helium too. It thus helps us in reducing the number of free 
parameters in our model with the escape fraction being given by a single free parameter $\eta_{\rm esc}$.

\item  {\it Self-consistent calculation of the temperature-density relation}: 
For calculations of the transmitted flux of the IGM (as would be observed in QSO absorption line experiments), 
it is usually assumed that the temperature-density relation follows a power-law form, $T \propto \Delta^{\gamma - 1}$. 
In this work, we solve the temperature evolution equation 
for fluid elements of different densities and thus
obtain the value of $\gamma$ at each redshift in a self-consistent manner
\cite{hg97}.

\item {\it Additional observational constraints}: In addition to the observations described in CF05, we use a few 
additional constraints to determine our free parameters. The most notable of these is the experiments related 
to the source counts at high redshifts \cite{bitf05}.  Three possible high redshift candidates have been identified 
by applying the J-dropout technique to the NICMOS HUDF; however the precise nature of these
three sources could not be confirmed.  Hence, \citeN{bitf05} concluded that the actual number of $z \approx 10$ 
sources in the NICMOS parallel fields must be three or fewer.

The number of sources above a redshift $z$ observed within a solid angle $\de \Omega$ in the flux range 
$[F_{\nu_0}, F_{\nu_0} + \de F_{\nu_0}]$ is 
\be
N_{F_{\nu_0}}(>z) = 
\f{\de N}{\de \Omega \de F_{\nu_0}}(F_{\nu_0}, z) = \int_z^{\infty}
\de z' \f{\de V}{\de z' \de \Omega} \f{\de n}{\de F_{\nu_0}} (F_{\nu_0}, z')
\e
where $\de V/\de z' \de \Omega$ denotes the comoving volume element per unit redshift per unit solid angle, and 
\be
\f{\de n}{\de F_{\nu_0}} (F_{\nu_0}, z') = \int_{z'}^{\infty}
\de z'' \f{\de M}{\de F_{\nu_0}}(F_{\nu_0}, t_{z'}-t_{z''})
\f{\de^2 n}{\de M \de z''}(M,z'')
\e
is the comoving number of objects at redshift $z'$ with observed flux within 
$[F_{\nu_0}, F_{\nu_0} + \de F_{\nu_0}]$.
The quantity $\de^2 n/\de M \de z''$ gives the formation rate of haloes of mass $M$, calculated using 
Press-Schechter formalism.  The flux is related to the mass of the halo $M$ by the relation
\be
F_{\nu_0} =
\f{\epsilon_{*} (\Omega_b/\Omega_m) M \int \de \nu' ~ l_{\nu'}(t_{z'}-t_{z''})~
{\rm e}^{-\tau_{\rm eff}(\nu_0,z=0,z')}}{4 \pi d_L^2(z') \Delta \nu_0}
\e
where $\epsilon_{*}$ is the star-forming efficiency of the 
population under consideration, $l_{\nu'}(t_{z'}-t_{z''})$ is template
luminosity per unit solar mass for the stellar population
of age $t_{z'}-t_{z''}$ (the time elapsed between the two redshifts), 
$d_L(z')$ is the luminosity distance and $\Delta \nu_0$ is the instrumental
bandwidth. The quantity $\tau_{\rm eff}(\nu_0,z=0,z')$ is the effective
optical depth at $\nu_0$ between $z'$ and $z = 0$, which can be calculated
self-consistently from the semi-analytical model given the 
density distribution. While calculating the source distribution, we
apply the same selection criteria as is used in the observational
analysis. For calculating the template luminosity $l_{\nu}$, we 
use stellar population models of \citeN{bc03} for PopII stars
and of \citeN{schaerer02} for PopIII stars.

We have also incorporated constraints from
the observed transmitted flux in the Ly$\beta$ region of the QSO absorption
spectra (in addition to Ly$\alpha$), setting more severe constraints on the background ionizing flux.

\end{itemize}

\subsection{Free parameters}

For the background cosmology, we use the best-fit parameters as given by the 3-year WMAP data \cite{sbd++06}, i.e., 
we assume a flat universe with total matter, vacuum, and baryonic densities in units of the
critical density of $\Omega_m = 0.24$, $\Omega_{\Lambda} = 0.76$, and
$\Omega_b h^2 = 0.022$, respectively, and a Hubble constant of $H_0 =
100\,h$ km s$^{-1}$ Mpc$^{-1}$, with $h=0.73$.  The parameters
defining the linear dark matter power spectrum are $\sigma_8=0.74$,
$n_s=0.95$, $\de n_s/\de \ln k =0$.
The reionization model, after improvements and modifications, contain 
just four free parameters. They are the 
star-forming efficiencies of PopII and PoIII stars ($\epsilon_{*,{\rm II}}$
and $\epsilon_{*,{\rm III}}$ respectively), $\eta_{\rm esc}$ related
to the escape fractions of the two stellar populations 
[see equations (\ref{eq:eta_esc})-(\ref{eq:f_esc_3})]
and 
the normalization of the mean
free path $\lambda_0$ [equation (\ref{eq:lambda_0})]. 
The main exercise of this paper
would be to find the range of parameter values which can match
all the observational data we have considered. A detailed exploration of the
parameter space using a statistical approach would be reported elsewhere.

\section{Results}

We now constrain the above free parameters and select the best-fit model as the one
that fits {\it simultaneously} all the available experimental data.
\subsection{The best-fit model}

\nocite{nchos04,songaila04,smih94,stle99}
\begin{figure*}
\rotatebox{270}{\resizebox{0.65\textwidth}{!}{\includegraphics{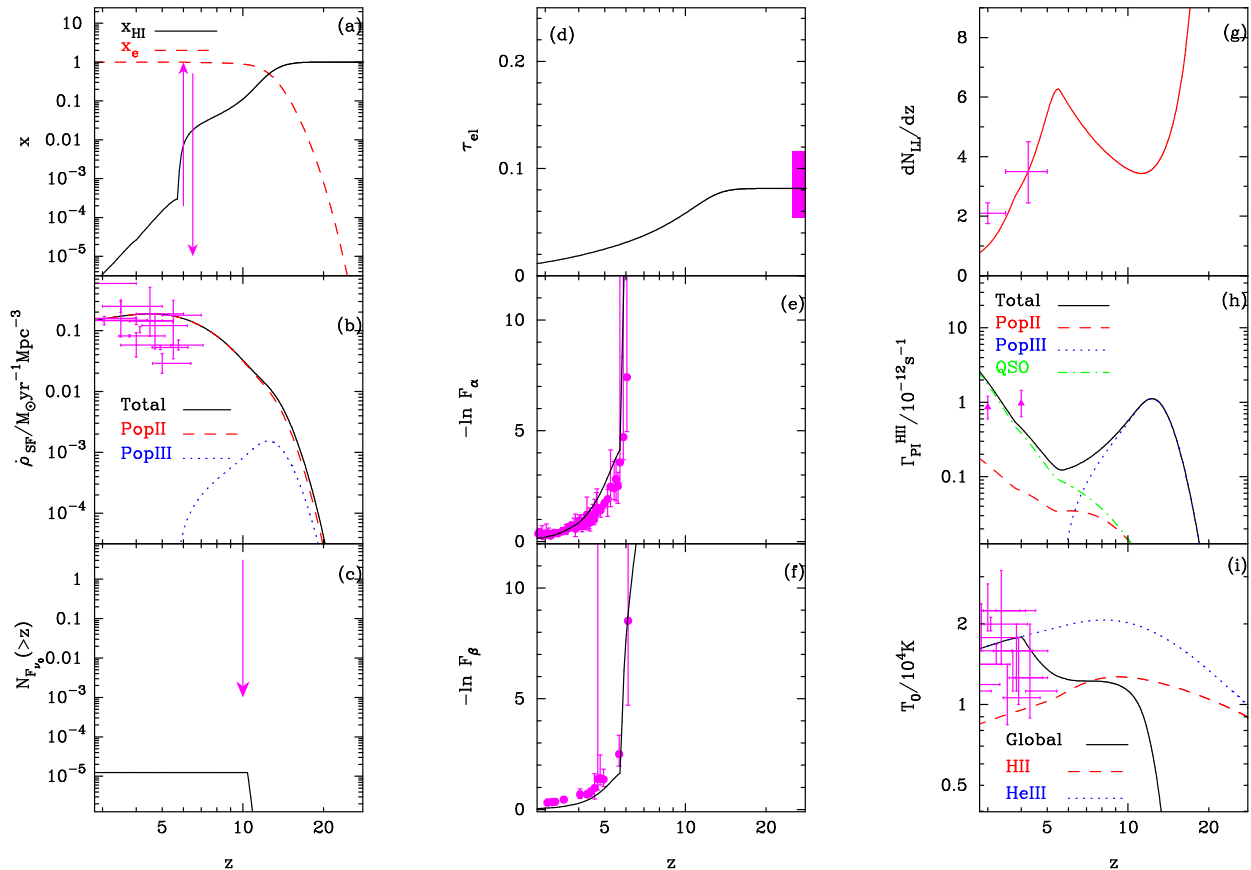}}}
\caption{Comparison of model predictions with observations for
the best-fit model with 
parameter values $\epsilon_{*,{\rm II}} = 0.2, 
\epsilon_{*,{\rm III}} = 0.07, f_{\rm esc, II} = 0.003, f_{\rm esc, III} = 
0.72$ (keeping in mind that $f_{\rm esc, II}$ and $f_{\rm esc, III}$
are not independent). The different panels indicate:
(a) The volume-averaged neutral hydrogen fraction $x_{\rm HI}$, with observational
lower limit from QSO absorption lines at $z =6$ and upper limit from
Ly$\alpha$ emitters at $z = 6.5$ (shown with arrows). In addition, the 
ionized fraction $x_e$ is shown by the dashed line.
(b) SFR for different stellar populations. The points with error-bars
indicate low-redshift observations taken from the compilation 
of  Nagamine et al. (2004) 
(c) The number of source
counts above a given redshift, 
with the observational upper limit from NICMOS HUDF shown by the
arrow. The contribution to the source count is zero at low 
redshifts because of the 
J-dropout selection criterion.
(d) Electron scattering optical depth, with  observational constraint from
WMAP 3-year data release. (e) Ly$\alpha$ effective optical
depth, with observed data points from Songaila (2004). 
(f) Ly$\beta$ effective optical
depth, with observed data points from Songaila (2004).
(g) Evolution of Lyman-limit systems, with observed data points from Storrie-Lombardi et al. (1994). (h) Photoionization
rates for hydrogen, with estimates from numerical simulations (shown
by points with error-bars; Bolton et al. 2005). (i) Temperature of the mean
density IGM, with observational estimates from Schaye et al. (1999).
}
\label{fig:bf}
\end{figure*}

The best-fit model is characterized by the parameter values
$\epsilon_{*,{\rm II}} = 0.2, \epsilon_{*,{\rm III}} = 0.07, f_{\rm esc, II} = 0.003, f_{\rm esc, III} = 
0.72$ (keeping in mind that $f_{\rm esc, II}$ and $f_{\rm esc, III}$ are not independent). The comparison 
between the best-fit model and different observations is shown in nine panels of Figure \ref{fig:bf}. 
Quite remarkably, the model matches a wide variety of observations by fitting only four parameters.

It is clear from the evolution of the neutral hydrogen fraction $x_{\rm HI}$ [Panel (a)] that 
current observations favour a model where reionization starts around $z = 15$ and is 90 per cent complete
by $z \approx 10$.  

The initial phase of reionization is driven by metal-free (PopIII) stars
which are capable of producing a large number of ionizing photons, as can be seen from the evolution of the 
photoionization rate of neutral hydrogen $\Gamma^{\rm HII}_{\rm PI}$ [Panel (h)]. 
The PopIII star-formation is severely quenched at $z \lesssim 10$ both because of the concomitant action of
both radiative and chemical feedback. 
In fact, chemical feedback allows PopIII star formation only in small
mass ($\lesssim 2 \times 10^8 M_{\odot}$) haloes, while radiative feedback tends to prohibit star formation in those, 
thus gradually terminating the PopIII stars.  The total mass of PopIII stars formed is $\Omega_{*, {\rm III}} \approx 10^{-6} $. As PopIII stars can no longer be formed, the progress of 
ionization fronts is limited, and hence the reionization extends and is completed only by $z \approx 6$. 
This evolution of $x_{\rm HI}$ is consistent both with high GP optical depths for Ly$\alpha$ [Panel (e)] and
Ly$\beta$ [Panel (f)] and with the constraints from Ly$\alpha$ emitters at $z \approx 6.5$ [Panel (a)].

At lower redshifts, the ionizing background seems to be dominated by QSOs as the PopII stars
have negligible escape fraction. Unlike CF05, the escape fractions of the PopII and PopIII stars 
in this work are determined by the same quantity $\eta_{\rm esc}$; hence it is {\it not} possible
to hike up $f_{\rm esc, II}$ without affecting $f_{\rm esc, III}$, in turn possibly violating some 
other observational constraints. At $z \approx 3-5$, our best-fit model is consistent with the observed
evolution of the SFR [Panel (b)], Lyman-limit systems [Panel (g)] and the temperature of the mean density IGM 
[Panel (i)]; finally, our estimates of $\Gamma^{\rm HII}_{\rm PI}$ are consistent with those obtained from numerical 
simulations of \cite{bhvs05} [Panel (h)].
Note that the SFR is almost always dominated by PopII stars [Panel (b)] as metal-free stars have          
a low star formation efficiency (3 times smaller than that of PopII stars); however, PopIII stars 
can still dominate the photoionization rate at high redshifts as they
produce larger number of photons per unit mass of stars formed.

The best-fit model produces negligible source counts at $z \approx 10$, which suggests that the three NICMOS
HUDF candidates  at $z\approx 10$ are probably spurious detections. 
Note that the main contribution to the source counts comes from the nebular and Ly$\alpha$ line emission 
of the PopIII stars.  The low value of source count is mainly determined by the required high value of 
$f_{\rm esc,III}$; since most of the photons escape the host halo, the amount of nebular and Ly$\alpha$ 
line emission is small, and hence no sources are above the detection threshold of the NICMOS experiments. 
In this sense, one can rule out this best-fit model if at least one of the sources observed in the NICMOS HUDF 
turns out be indeed at $z \approx 10$. We shall discuss this possibility later. 
Before addressing other issues, it might be worthwhile mentioning the 
ionization history of doubly-ionized helium. We find that the escape fraction for photons with energies 
above 54.4 eV is not very high for PopIII stars, and hence the propagation
of doubly-ionized helium fronts is not very efficient at high redshifts. 
The complete reionization occurs only around $z \approx 3.5$ because of
QSOs.

\subsection{Variants of the best-fit model}

In spite of the success of our best-fit model in fitting observations, it is instructive to study some 
of its variants. In particular, we ask some interesting questions and try to find what the current data imply:

(i) Is it possible to fit the data with reionization at higher redshifts? By increasing $\epsilon_{*,{\rm III}}$ 
one can force an earlier start of the reionization process. For example, a model with parameter values $\epsilon_{*,{\rm II}} = 0.2, 
\epsilon_{*,{\rm III}} = 0.2, f_{\rm esc, II} = 0.006, f_{\rm esc, III} = 0.9$ gives a nearly equal good fit to the data.
In this model, reionization starts much earlier and  the IGM is 95 (99) per cent ionized by $z \approx 10 (8)$.
However, as in the best-fit model, the contribution of the PopIII stars to the ionizing flux 
decreases because of feedback and hence the reionization is extended till $z \approx 6$. Interestingly, the most 
severe constraint on the early reionization scenarios comes from Ly$\beta$ observations at $z \approx 6$ which rule 
out high values of $\epsilon_{*,{\rm III}} f_{\rm esc, III}$.

(ii) What if one or some of the candidates in the NICMOS HUDF do turn out to be valid $z \approx 10$ sources? 
Then the  best-fit model above could be ruled out as it predicts negligible source counts at $z > 10$.
However, there are parameters within 2$\sigma$ of the best-fit value which predict high number of sources at 
$z \approx 10$.  For example, a model with parameter values $\epsilon_{*,{\rm II}} = 0.2, 
\epsilon_{*,{\rm III}} = 0.4, f_{\rm esc, II} = 0.0, f_{\rm esc, III} = 0.36$ predicts $\approx 2.6$
sources at $z \approx 10$.  The condition for producing high source counts is that
the escape fraction of PopIII stars should be low, which in turn means
that the ionizing flux is lower and reionization would be delayed.
Such models tend to overpredict the Ly$\alpha$ optical depth at $z \approx 4-5$ and thus the 
parameter space is severely constrained.  In other words,  if one observes sources at high redshifts, 
a very small parameter space would be allowed, and we would possibly be able to identify uniquely how 
reionization occurred.

(iii) Do we still require a prolonged epoch of metal-free star formation to explain observations? 
In fact, one can explain the low redshift SFR constraints and the WMAP $\tau_{\rm el}$ without PopIII 
stars provided $\epsilon_{*,{\rm II}} = 0.2$ and $f_{\rm esc} > 0.07$. However, such a high value of 
escape fraction violates the Ly$\alpha$ and Ly$\beta$ optical depths at $z \approx 3-4$; 
the GP optical depth measurements require that $f_{\rm esc} < 0.05$, which when combined with the WMAP
$\tau_{\rm el}$, gives $\epsilon_{*,{\rm II}} > 0.02$. Thus a combination of low redshift SFR, $\tau_{\rm el}$ 
and GP optical depth constraints imply that we still do require a non-zero contribution from metal-free
PopIII stars, albeit with a small star-forming efficiency.

\section{Summary}

We have extended the self-consistent model of CF05 incorporating key additional physical processes 
and compared the model predictions with a variety of data sets. Our formalism now includes
the inhomogeneous IGM density distribution, three different classes of ionizing photon sources 
(metal-free PopIII stars, PopII stars and QSOs), chemical feedback inhibiting formation of PopIII 
stars in metal-enriched haloes, and radiative feedback preventing the formation of stars in galaxies 
below a certain circular velocity threshold. Our model is able to: (i) predict the star formation/emissivity 
history of sources and the number of sources above a given flux detection threshold at various redshifts, 
(ii) follow the evolution of H and He reionization and of the intergalactic gas temperature, and (iii) 
yield a number of additional predictions involving directly observable quantities.

By constraining the model free parameters with the available experimental data 
we have found a best-fit  model and a set of allowed parameter values which 
matches very well all the available observations. From this analysis, 
an updated
reionization scenario, which also takes into account the 3-year WMAP data, emerges: 

\begin{itemize}
\item Hydrogen reionization starts around $z \approx 15$ driven by
metal-free (PopIII) stars, and it is 90 per cent complete by $z \approx 10$. 
The contribution of PopIII stars decrease below this redshift because of the combined action 
of radiative and chemical feedback. As a result, reionization is extended considerably completing
only at $z \approx 6$.

\item Scenarios in which reionization is completed much earlier are ruled out by a combination
of constraints from $\tau_{\rm el}$, the NICMOS source counts at $z \approx 10$ and the Ly$\beta$ 
optical depths at $z \approx 6$.

\item The combination of $\tau_{\rm el}$ and GP optical depth constraints require non-zero 
contribution from metal-free stars with a normal Salpeter IMF. Non-inclusion of PopIII 
stars would require a relatively larger ionization flux from PopII stars to match the WMAP 
$\tau_{\rm el}$, which would then violate the GP optical depth constraints.

\end{itemize}

\section*{Acknowledgments}

A special thanks goes to R. Schneider for providing us with the data on chemical feedback.
We would like to thank B. Ciardi, M. Mapelli, S. Gallerani, and R. Salvaterra for enlightening discussions.

\bibliography{mnrasmnemonic,astropap-mod,reionization}

\begin{thebibliography}{}

\bibitem[\protect\citeauthoryear{{Bolton} et~al.}{{Bolton}
  et~al.}{2005}]{bhvs05}
{Bolton} J.~S., {Haehnelt} M.~G., {Viel} M.,  {Springel} V., 2005, MNRAS, 357,
  1178

\bibitem[\protect\citeauthoryear{{Bouwens} et~al.}{{Bouwens}
  et~al.}{2005}]{bitf05}
{Bouwens} R.~J., {Illingworth} G.~D., {Thompson} R.~I.,  {Franx} M., 2005, ApJ,
  624, L5

\bibitem[\protect\citeauthoryear{{Bruzual} \& {Charlot}}{{Bruzual} \&
  {Charlot}}{2003}]{bc03}
{Bruzual} G.,  {Charlot} S., 2003, MNRAS, 344, 1000

\bibitem[\protect\citeauthoryear{{Choudhury} \& {Ferrara}}{{Choudhury} \&
  {Ferrara}}{2005}]{cf05}
{Choudhury} T.~R.,  {Ferrara} A., 2005, MNRAS, 361, 577

\bibitem[\protect\citeauthoryear{{Ciardi}, {Ferrara}, \& {White}}{{Ciardi}
  et~al.}{2003}]{cfw03}
{Ciardi} B., {Ferrara} A.,  {White} S.~D.~M., 2003, MNRAS, 344, L7

\bibitem[\protect\citeauthoryear{{Fan} et~al.}{{Fan} et~al.}{2001}]{fnl+01}
{Fan} X. et~al., 2001, AJ, 122, 2833

\bibitem[\protect\citeauthoryear{{Fan} et~al.}{{Fan} et~al.}{2005}]{fsb++05}
{Fan} X. et~al., 2005, Preprint: astro-ph/0512082

\bibitem[\protect\citeauthoryear{{Fan} et~al.}{{Fan} et~al.}{2003}]{fss++03}
{Fan} X. et~al., 2003, AJ, 125, 1649

\bibitem[\protect\citeauthoryear{{Gallerani}, {Choudhury}, \&
  {Ferrara}}{{Gallerani} et~al.}{2005}]{gcf05}
{Gallerani} S., {Choudhury} T.~R.,  {Ferrara} A., 2005, Preprint:
  astro-ph/0512129

\bibitem[\protect\citeauthoryear{Haiman \& Bryan}{Haiman \& Bryan}{2006}]{hb06}
Haiman Z.,  Bryan G.~L., 2006, Preprint: astro-ph/0603541

\bibitem[\protect\citeauthoryear{{Hinshaw} et~al.}{{Hinshaw}
  et~al.}{2006}]{hnb++06}
{Hinshaw} G. et~al., 2006, Preprint: astro-ph/0603451

\bibitem[\protect\citeauthoryear{{Hui} \& {Gnedin}}{{Hui} \&
  {Gnedin}}{1997}]{hg97}
{Hui} L.,  {Gnedin} N.~Y., 1997, MNRAS, 292, 27

\bibitem[\protect\citeauthoryear{{Kogut} et~al.}{{Kogut}
  et~al.}{2003}]{ksb++03}
{Kogut} A. et~al., 2003, ApJS, 148, 161

\bibitem[\protect\citeauthoryear{{Miralda-Escud{\' e}}, {Haehnelt}, \&
  {Rees}}{{Miralda-Escud{\' e}} et~al.}{2000}]{mhr00}
{Miralda-Escud{\' e}} J., {Haehnelt} M.,  {Rees} M.~J., 2000, ApJ, 530, 1

\bibitem[\protect\citeauthoryear{{Nagamine} et~al.}{{Nagamine}
  et~al.}{2004}]{nchos04}
{Nagamine} K., {Cen} R., {Hernquist} L., {Ostriker} J.~P.,  {Springel} V.,
  2004, ApJ, 610, 45

\bibitem[\protect\citeauthoryear{{Page} et~al.}{{Page} et~al.}{2006}]{phk++06}
{Page} L. et~al., 2006, Preprint: astro-ph/0603450

\bibitem[\protect\citeauthoryear{{Salvaterra} \& {Ferrara}}{{Salvaterra} \&
  {Ferrara}}{2005}]{sf05}
{Salvaterra} R.,  {Ferrara} A., 2005, Preprint: astro-ph/0509338

\bibitem[\protect\citeauthoryear{{Schaerer}}{{Schaerer}}{2002}]{schaerer02}
{Schaerer} D., 2002, A\&A, 382, 28

\bibitem[\protect\citeauthoryear{{Schaye} et~al.}{{Schaye}
  et~al.}{1999}]{stle99}
{Schaye} J., {Theuns} T., {Leonard} A.,  {Efstathiou} G., 1999, MNRAS, 310, 57

\bibitem[\protect\citeauthoryear{{Schneider} et~al.}{{Schneider}
  et~al.}{2002}]{sfno02}
{Schneider} R., {Ferrara} A., {Natarajan} P.,  {Omukai} K., 2002, ApJ, 571, 30

\bibitem[\protect\citeauthoryear{{Schneider} et~al.}{{Schneider}
  et~al.}{2003}]{sfsob03}
{Schneider} R., {Ferrara} A., {Salvaterra} R., {Omukai} K.,  {Bromm} V., 2003,
  Nat, 422, 869

\bibitem[\protect\citeauthoryear{{Schneider} et~al.}{{Schneider}
  et~al.}{2005}]{ssfc05}
{Schneider} R., {Salvaterra} R., {Ferrara} A.,  {Ciardi} B., 2005, Preprint:
  astro-ph/0510685

\bibitem[\protect\citeauthoryear{{Songaila}}{{Songaila}}{2004}]{songaila04}
{Songaila} A., 2004, AJ, 127, 2598

\bibitem[\protect\citeauthoryear{{Spergel} et~al.}{{Spergel}
  et~al.}{2006}]{sbd++06}
{Spergel} D.~N. et~al., 2006, Preprint: astro-ph/0603449

\bibitem[\protect\citeauthoryear{{Spergel} et~al.}{{Spergel}
  et~al.}{2003}]{svp++03}
{Spergel} D.~N. et~al., 2003, ApJS, 148, 175

\bibitem[\protect\citeauthoryear{{Storrie-Lombardi} et~al.}{{Storrie-Lombardi}
  et~al.}{1994}]{smih94}
{Storrie-Lombardi} L.~J., {McMahon} R.~G., {Irwin} M.~J.,  {Hazard} C., 1994,
  ApJ, 427, L13

\end{thebibliography}
\bibliographystyle{mnras}

\end{document}